\documentclass[10pt]{article}
\usepackage{amsmath}
\usepackage{amssymb}
\usepackage{natbib}

\begin{document}
\bibliographystyle{plainnat}
\setcitestyle{numbers,square}

\title{\normalsize{QUASI-RADIAL MODES OF PULSATING NEUTRON STARS: \\
                   NUMERICAL RESULTS FOR GENERAL-RELATIVISTIC    \\
                   RIGIDLY ROTATING POLYTROPIC MODELS}
                  }

\author{Vassilis Geroyannis$^1$, Eleftheria Tzelati$^2$ \\
        $^{1,2}$Department of Physics, University of Patras, Greece \\
        $^1$vgeroyan@upatras.gr, $^2$etzelati@physics.upatras.gr}

\maketitle

\begin{abstract}
In this paper we compute general-relativistic polytropic models simulating rigidly rotating, pulsating neutron stars. These relativistic compact objects, with a radius of $\sim 10 \, \mathrm{km}$ and mass between $\sim 1.4$ and $3.2$ solar masses, are closely related to pulsars. We emphasize on computing the change in the pulsation eigenfrequencies owing to a rigid rotation, which, in turn, is a decisive issue for studying stability of such objects. In our computations, we keep rotational perturbation terms of up to second order in the angular velocity.  
\end{abstract}

\section{Introduction}
According to ``Hartle's perturbation method'' (also called ``Hartle--Thorne perturbation method''; \citep{4H67}, \citep{5HT68}), we treat rotating relativistic neutron stars as perturbative solutions of the Einstein's field equations, which describe a static spherically symmetric relativistic model. Our aim here is to compute, based on the work of Hartle et al. \citep{13HTC72} and also of Hartle and Friedmann \citep{6HF75}, the zeroth- and second-order eigenfrequencies of the lowest three modes of radial pulsation for general-relativistic polytropic models. The zeroth-order eigenfrequencies, $[\sigma^2]^{(0)}$, are the eigenfrequencies of the nonrotating model, while the second-order ones, $[\sigma^2]^{(2)}$, are the rotation-induced changes in the eigenfrequencies. 

In this study, we use ``gravitational units'' (abbreviated ``gu''; see e.g. \citep{7GK08} and \citep{8GS11}), also called ``geometrized units''. In these units the speed of light, $c$, and the gravitational constant, $G$, are equal to unity, $c_\mathrm{gu} = G_\mathrm{gu} = 1$, and the length is the unique ``base unit'', measured in centimeters (abbreviated ``cm''). On the other hand, the well-known ``centimeter--gram--second units'' (abbreviated ``cgs'') have three base units: the length measured in cm, the mass measured in grams (abbreviated ``g''), and the time measured in seconds (abbreviated ``s'').

\section{The Nonrotating Model}
\label{NRM}
The Schwarzschild metric for a nonrotating spherical object, expressed in polar coordinates, is given by the relation (\citep{4H67}, Equation (25))
\begin{equation}
ds^2=-e^{\nu}dt^2+e^{\lambda}dr^2+r^2(d\theta^2+\sin^2\theta d\phi^2),
\end{equation}
where $\nu$ and $\lambda$ are metric functions of $r$. The exponential functions $e^{\nu}$ and $e^{\lambda}$ represent the fluctuations in the time rate flow and the divergence from the Euclidean geometry, respectively. The gravitational potential $\Phi$ is usually defined as (see e.g. \citep{7GK08}, Equation (2)) 
\begin{equation}
\Phi=\frac{\nu}{2},
\end{equation}
and the metric function $\lambda$ by (see e.g. \citep{7GK08}, Equation (5))
\begin{equation}
e^{\lambda} = \left( 1-\frac{2m}{r} \right)^{-1},
\end{equation}
where $m(r)$ is the mass-energy of the star.

A relativistic neutron star obeys the Tolman--Oppenheimer--Volkoff (TOV) equations of (i)  hydrostatic equilibrium (\citep{4H67}, Eq. (28))
\begin{equation}
\frac{dP}{dr} = - \, \, \frac{(E+P)(m + 4 \pi \, r^3 \, P)}{r(r-2m)},
\label{TOVP}
\end{equation}
and (ii) mass--energy (\citep{4H67}, Eq. (29a))
\begin{equation}
\frac{dm}{dr}=4\pi r^2 E,
\label{TOVM}
\end{equation}
fulfilling the initial conditions
\begin{equation}
P(0) = P_c = P(E_c), \ \ E(0) = E_c, \  \ m(0) = m_c = 0,
\label{TOVIC}
\end{equation}
where $P(r)$ is the pressure and $E(r)$ the mass-energy density.

To treat numerically the system (\ref{TOVP})--(\ref{TOVM}), we need an equation relating $P$ to $E$, $P = P(E)$, that is, an ``equation of state'' (EOS). In this study, we consider the polytropic EOS (see e.g. \citep{9CST94}, Equation (3); see also \citep{10T65}, Section II)
\begin{equation}
P = K \, \varrho^{1 + (1/n)}.
\label{poleos}
\end{equation}
The parameters $K$ and $n$ are the so-called "polyropic constant" and "polytropic index", respectively. The so-called ``adiabatic index'', $\Gamma$, is defined as (cf. \citep{13HTC72}, Equation~(2.3))
\begin{equation}
\Gamma = \, \frac{E+P}{P} \, \left(\frac{dP}{dE} \right)_\mathrm{constant \, entropy},
\label{Gamma}
\end{equation}
and, in general, is associated with perturbations about equilibrium under constant entropy. On the other hand, the so-called ``adiabatic index associated with the equation of state'', $\gamma$, is defined as (cf. \citep{13HTC72}, Equation~(3.12))
\begin{equation}
\gamma = \, \frac{E+P}{P} \, \left(\frac{dP}{dE} \right)_\mathrm{equation \, of \, state}.
\label{gamma}
\end{equation}
In the polytropic EOS~(\ref{poleos}), $\varrho(r)$ is the rest-mass density, related to the mass-energy density via the equation (\citep{8GS11}, Equation (9))
\begin{equation}
E = \varrho + n \, P.
\end{equation}

To solve the system (\ref{TOVP})--(\ref{TOVM}), we write the two equations as (\citep{8GS11}, Equations (8) and (9), respectively)
\begin{equation}
\frac{d\varrho}{dr}=-\frac{[\varrho+(1+n)P](m+4\pi r^3 P)}{r(r-2m)(dP/d\varrho)}
=-\frac{[\varrho+K(1+n)\varrho^{\Gamma}](m+4K\pi r^3\varrho^{\Gamma})}{K\Gamma r(r-2m)\varrho^{1/n}},
\end{equation}
\begin{equation}
\frac{dm}{dr}=4\pi r^2(\varrho+nP)=4\pi r^2(\varrho+Kn\varrho^{\Gamma}).
\end{equation}
The initial conditions are now 
\begin{equation}
\varrho(0) = \varrho_c, \ \ m(0)=0.
\label{MODIC}
\end{equation}

To complete the study of the nonrotating model, we need to solve the differential equation for the gravitational potential $\Phi$ (\citep{4H67}, Equation (29b))
\begin{equation}
\frac{d\Phi}{dr}= - \, \, \frac{1}{E+P} \, \frac{dP}{dr} =
\frac{m + 4 \pi K r^3 \varrho^{\,\Gamma}}{r(r-2m)},
\end{equation}
obeying the boundary condition at the surface of the star (cf. \citep{8GS11}, Equation~(30)) 
\begin{equation}
\Phi_R = \, \frac{1}{2} \, \ln \left( 1-\frac{2M}{R} \right).
\end{equation}

\section{Rigid Rotation}
\label{RR}
When a star is rotating, its shape deviates from sphericity. Furthermore, when the star is in equilibrium state,  there is a balance between the pressure forces, the gravitational forces, and the centrifugal forces. Assuming that the star is rotating rigidly and slowly (see however the starting remark in Section~\ref{NRD}), the perturbed metric is given by (\citep{5HT68}, Equation~(4))
\begin{equation}
\begin{aligned}
ds^2 &= -e^{\nu} \left[ 1 + 2(h_0+h_2P_2) \right] dt^2 + 
\frac{1+2(m_0+m_2P_2)/(r-2m)}{1-2m/r} \, dr^2 \\
&+ r^2 \left[ 1 + 2(\upsilon_2-h_2)P_2 \right] 
\left[ d\theta^2+sin^2\theta(d\phi-\omega dt)^2 \right] + \mathcal{O}(\Omega^3),
\end{aligned}
\label{pm}
\end{equation}
where $P_l=P_l(\cos \theta)$ are Legendre polynomials; the perturbation functions $m_0, \, h_0, \, m_2, \, h_2,$ and $\upsilon_2$ are radial functions proportional to $\Omega^2$;
the centrifugal forces depend on the angular velocity $\Omega$ relative to a distant observer, as well as on the angular velocity $\varpi$ relative to the local inertial frame, these two angular velocities connected via the relation (\citep{5HT68}, Equation~(6))
\begin{equation}
\varpi = \Omega - \omega,
\end{equation}
where $\omega$ is the angular velocity of the local inertial frame.
To calculate $\varpi$, we solve the so-called ``frame-dragging equation'' (\citep{5HT68}, Eq. (9))
\begin{equation}
\frac{1}{r^4} \, \frac{d}{dr} \left( r^4 \, j \, 
    \frac{d\varpi}{dr} \right) + \, \frac{4}{r} \, \frac{dj}{dr}\, \varpi = 0,
\label{fd}
\end{equation}
where (\citep{5HT68}, Equation~(10))
\begin{equation}
j = e^{-\Phi} \, \left( 1 - \frac{2 \, m}{r} \right)^{1/2}.
\end{equation}
Outside the star, $\varpi$ takes the form (\citep{5HT68}, Equation~(13b))
\begin{equation}
\varpi = \Omega - \frac{2 \, J}{r^3},
\label{varpi}
\end{equation}
where $J$ is the total angular momentum of the star, given by (\citep{5HT68}, Equation~(13a))
\begin{equation}
J = \, \frac{1}{6} \, R^4 \, \left( \frac{d\varpi}{dr} \right)_{r=R},
\label{J}
\end{equation}
where $R$ is the radius of the star.

To solve Equation~(\ref{fd}), we integrate from the center of the star outwards, imposing the initial conditions 
\begin{equation}
\varpi = \varpi_c, \ \ \ \ \ \ \frac{d\varpi}{dr} = 0, 
\end{equation}
where the constant $\varpi_c$ is chosen arbitrarily. Since, due to Equation~(\ref{varpi}), the angular velocity $\Omega_\mathrm{arb}$ corresponding to the arbitrarily chosen initial value $\varpi_c$ is equal to 
\begin{equation}
\Omega_\mathrm{arb} = \varpi(R) + \, \frac{2J}{R^3},
\end{equation}
and since, in general, the prescribed angular velocity $\Omega$ is different to $\Omega_\mathrm{arb}$, we must rescale the solution $\varpi(r)$ of Equation~(\ref{fd}) in order for the prescribed value $\Omega$ to be applied to our model (for details on this matter, see \citep{7GK08}, Section 5.2),
\begin{equation}
\varpi_\mathrm{new}(r) = \, \frac{\Omega}{\Omega_\mathrm{arb}} \, \, \varpi(r).
\end{equation}

\section{Spherical Deformation}
\label{SD}
The spherical deformations due to rigid rotation are the mass perturbation function, $m_0$,  and the pressure perturbation function, $p_0$. They are calculated by integrating the ``$l=0$ equations of hydrostatic equilibrium'' (\citep{4H67}, Section~VII), which become after a long algebra (\citep{5HT68}, Equations~(15a) and (15b), respectively)
\begin{equation}
\frac{dm_0}{dr}=4\pi r^2 \frac{dE}{dP}(E+P)p_0+\frac{1}{12}j^2r^4
\left( \frac{d\varpi}{dr} \right)^2-\frac{1}{3}r^3 \, \frac{dj^2}{dr} \, \varpi^2,
\label{m0}
\end{equation}

\begin{equation}
\begin{aligned}
\frac{dp_0}{dr} &= \, - \, \frac{m_0(1+8 \pi \, r^2 \, P)}{(r-2m)^2} \, - \, \frac{4\pi \, (E+P) \, r^2}{r-2m} \, p_0 \\ 
&+ \, \frac{1}{12} \, \frac{r^4 \, j^2}{r-2m} \left(\frac{d\varpi}{dr} \right)^2 \,
 + \, \frac{1}{3} \, \frac{d}{dr} \left( \frac{r^3 \, j^2 \, \varpi^2}{r-2m} \right),
\end{aligned}
\end{equation}
with initial conditions $m_0=0$ and $p_0=0$ at $r = 0$.

The perturbation function $h_0$ (to be used in Section 5), involved in the metric~(\ref{pm}), is defined outside the star as (\citep{5HT68}, Equation~(17a))
\begin{equation}
h_0 = - \, \frac{\delta M}{r-2 \, M} + \, \frac{J^2}{r^3(r-2 \, M)}
\end{equation}
and inside the star as (\citep{5HT68}, Equation~(17b))
\begin{equation}
h_0 = -p_0 + \, \frac{1}{3} \, r^2 \, e^{-\nu}\varpi^2 + h_{0\mathrm{c}}.
\end{equation}
The constant $h_{0\mathrm{c}}$ is determined so that $h_0$ be continuous across the surface (see e.g. \citep{7GK08}, Equation~(47))
\begin{equation}
h_{0c}=-\frac{1}{R-2M} \, m_0(R) + p_0(R)- \, \frac{1}{3} \, R^2e^{-2\Phi(R)} \, \varpi^2(R).
\end{equation}
In these equations, $M$ is the mass-energy of the star (also called gravitational mass); and $\delta M$ is the increase in the gravitational mass due to spherical deformation, given by (\citep{7GK08}, Equation~(43))
\begin{equation}
\delta M = m_0(R)+ \, \frac{J^2}{R^3}. 
\end{equation}

\section{Quasi-Radial Pulsation}
\label{RO}
When a star is rotating rigidly with a small angular velocity $\Omega$ (see however the starting remark in Section~\ref{NRD}), the squared eigenfrequencies $\sigma^2$ of its pulsation modes can be expanded in powers of $\Omega$ (\citep{6HF75}, Equation~(2.1)),
\begin{equation}
\sigma^2 = [\sigma^2]^{(0)} + [\sigma^2]^{(2)} + \dots 
\label{wdistort}
\end{equation}
where the superscript ``$(0)$'' denotes terms of zeroth order in $\Omega$, and the superscript ``$(2)$'' terms of second order in $\Omega$.
The aim of the present study is to compute the zeroth- and the second-order eigenfrequencies of the lowest three pulsation modes.

Concerning the zeroth-order eigenfrequencies, we begin with the so-called ``Chandrasekhar operator'' (\citep{13HTC72}, Equation~(4.6b)), $\mathcal{L}[U]$, applied to the so-called ``displacement function'' (\citep{13HTC72}, Equations~(4.2a,\,b,\,c,\,d)), $U$, and set equal to zero, 
\begin{equation}
\mathcal{L}[U] = \mathcal{W} \, [\sigma^2]^{(0)} \, U +
                 ( \mathcal{A} \, U' )' +
                 \mathcal{F}_1 (-\mathcal{F}_2 - \mathcal{F}_3 + \mathcal{F}_4)\,U = 0,  
\label{LU}
\end{equation}
where primes denote differentiation with respect to $r$, and
\begin{equation}
\mathcal{F}_1 = e^{(3\nu + \lambda)/2},
\end{equation}
\begin{equation}
\mathcal{F}_2 = \, \frac{4 \, P'}{r^3},
\end{equation}
\begin{equation}
\mathcal{F}_3 = 8 \, \pi \, e^\lambda \, \frac{P(E+P)}{r^2},
\end{equation}
\begin{equation}
\mathcal{F}_4 = \, \frac{(P')^2}{(E+P) \, r^2},
\end{equation}
\begin{equation}
\mathcal{A} = \, \frac{\mathcal{F}_1 \, \Gamma \, P}{r^2},
\end{equation}
\begin{equation}
\mathcal{W} = \, \frac{e^{(\nu + 3\lambda)/2} \, (E+P)}{r^2}.
\end{equation}
The second-order differential equation~(\ref{LU}), subject to the initial condition (\citep{13HTC72}, Equation~(4.2e))
\begin{equation}
U = \alpha \, r^3 \ \ \ \mathrm{near} \ \ r = 0
\label{SLIC}
\end{equation}
(without loss of generality, we can take $\alpha=1$) and to the boundary condition 
\begin{equation}
\Gamma \, P \, U' = 0 \ \ \ \mathrm{at \, the \, surface} \ \ r = R,
\label{SLBC}
\end{equation}
establishes a Sturm-Liouville (SL) boundary value problem with eigenvalues the pulsation eigenfrequencies  $[\sigma^2]^{(0)}$. Equation~(\ref{LU}) can be put into the so-called ``SL form'' (cf. \citep{14KR01}, Equation~(14))
\begin{equation}
(\mathcal{A} \, U')' + 
 \left(\mathcal{Q} + \mathcal{W} \, [\sigma^2]^{(0)}\right) U = 0,
\label{SL}
\end{equation}
where
\begin{equation}
\mathcal{Q} = \mathcal{F}_1 \left(-\mathcal{F}_2 - \mathcal{F}_3 + \mathcal{F}_4\right).
\end{equation} 
The SL form~(\ref{SL}) can be transformed into a system of two first-order differential equations (cf. \citep{14KR01}, Equations~(18) and (19)),
\begin{equation}
U' = \upsilon, \ \ \ \ \ \ 
\upsilon' = \, - \, \frac{\mathcal{A}' \, \upsilon + 
\left( \mathcal{Q} + \mathcal{W} \, [\sigma^2]^{(0)} \right) U}{\mathcal{A}}, 
\end{equation}
subject to the initial conditions $U = r^3$ and $\upsilon = 3 r^2$ near $r = 0$, and to the boundary condition $\Gamma \, P \, \upsilon = \Gamma \, P \, U' = 0$ at the surface (Equation~(\ref{SLBC})).

To compute the eigenvalue(s) $[\sigma^2]^{(0)}$, we work as follows. We start the numerical integration for a trial value  $\sigma^2$ and initial conditions as above. We integrate towards the surface and then check if the resulting solution $U$ satisfies the boundary condition $\Gamma \, P \, U(R)' = 0$. From the point of view of numerical analysis, this boundary condition can be treated as an algebraic equation of the form $f\left( \sigma^2 \right) = 0$; thus our task, to compute the root(s) $[\sigma^2]^{(0)}$ of this equation, can be achieved by a standard numerical method (e.g. the bisection method).  

The second-order eigenfrequencies $[\sigma^2]^{(2)}$ are computed by the relation (\citep{13HTC72}, Equation~(4.8))
\begin{equation}
[\sigma^2]^{(2)} = \, 
\frac{\int_0^R \left[ e^{\nu+\lambda/2} \, U(r) \, \mathfrak{D}(r) \right] dr}
     {\int_0^R \left[ \mathcal{W}(r) \, U^2(r) \right] dr}.
\end{equation}
The driving term $\mathfrak{D}(r)$ is defined as (cf. \citep{13HTC72}, Table~4)
\begin{equation}
\begin{aligned}
\mathfrak{D}= & \, \, U' \times 
\Biggl\{ m_0 \, r^{-4} \, e^{2\lambda+\Phi} \, \Gamma \left( E+P \right) + \mathcal{T}_0 + \frac{2}{3} \, \varpi^2 r^{-1} e^{\lambda-\Phi} \, \Gamma \, P \times \mathcal{T}_1 \Biggr\} \ \ + \\
& \, \, U \times \Biggl\{ m_0 \, r^{-5} \, e^{3\lambda+\Phi} \times \mathcal{S}_0 + 2  h_0 \, r^{-2} \, (E+P) \, e^{\lambda-\Phi} \, [\sigma^2]^{(0)} \ \ + \\
&p_0 \, \left( E+P \right) \, r^{-4} \, e^{2\lambda+\Phi} \times \mathcal{S}_1 + 4\varpi\varpi' \, r^{-1} e^{-\Phi} \left( E+P+\frac{1}{3}\,\Gamma P \right)  \ \ + \\
&\frac{2}{3} \, \varpi'^2 \, e^{\lambda-\Phi} \times \mathcal{S}_2 + \frac{2}{3} \, \varpi^2 \, r^{-2}\,  e^{\lambda-\Phi} \times \mathcal{S}_3 \Biggr\},
\end{aligned}
\end{equation} 
where, in turn,
\begin{equation}
\begin{aligned}
\mathcal{T}_0 = & \, \Biggl[ \frac{1}{2} \, p_0 \, \Gamma \, \left( E+P \right) r^{-3} e^{\lambda+\Phi} \ \ \times \Biggr. \\
& \left( \frac{E+P}{\gamma P}-\frac{E}{P} \right) \left( 1-e^{-\lambda} \right) -\frac{2}{3} \, \varpi \, \varpi' \, e^{-\Phi} \, \biggl( \Gamma \left( E+P \right) \ \ + \biggr. \\
& \Biggl. \biggl. \frac{2\,(\Gamma P)^2}{E+P} \biggr) 
-\frac{1}{12} \, \varpi'^2 \, re^{-\Phi} \, \Gamma \left( E+P \right) \Biggr] 
\end{aligned}
\end{equation}

\begin{equation}
\begin{aligned}
\mathcal{T}_1 = & \left[ -\frac{1}{2} \left( 3e^{-\lambda}-1 \right) \frac{E+P}{P}+\frac{1}{2} \, \frac{\Gamma P}{E+P} \left( 1-5e^{-\lambda} \right) \ \ +  \right. \\
& \frac{1}{2} \, \Gamma \left( 1-e^{-\lambda} \right) \left( 1-\frac{1}{\gamma} \right) +4\pi r^2 \Gamma P \left( 1-\frac{1}{\gamma}+\frac{P}{E+P} \right) \ \ -  \\
& \left. re^{-\lambda} \frac{\Gamma'P}{E+P} \right],
\end{aligned}
\end{equation}

\begin{equation}
\begin{aligned}
\mathcal{S}_0 = & \Biggl[ \Gamma(E+P) \biggl[ -\frac{1}{2} \left( 1-e^{-\lambda} \right) \biggr. \Biggr. \ \ + \\
& \biggl. 4\pi r^2P \left( 1+2e^{-\lambda} \right) +64\pi^2r^4P^2 \biggr] + (E+P+\Gamma P) \biggl[ -1-3e^{-\lambda} \biggr. \ \ - \\
& \biggl. 16\pi r^2P \left( 1 + \frac{1}{2} \ e^{-\lambda} \right) -64\pi^2r^4P^2 \biggr] +re^{-\lambda} \Gamma' P \left( 1+8\pi r^2P \right) \ \ - \\
& \Biggl. 2(E+P)[\sigma^{(0)}]^2r^2e^{-\lambda-\nu} \Biggr],
\end{aligned}
\end{equation}

\begin{equation}
\begin{aligned}
\mathcal{S}_1 = &  \biggl( \frac{E+P}{\gamma P}-\frac{E}{P} \biggr) \biggl[ -[\sigma^{(0)}]^2r^2 e^{\lambda-\nu} \ \ - \biggr. \\
& \biggl. \frac{1}{4} \left( 1-e^{-\lambda} \right) \left( 1+7e^{-\lambda} \right) \biggr] +4\pi \Gamma' Pr^3e^{-\lambda}-2\pi Pr^2 \biggl[ (1+e^{-\lambda})(2+\Gamma) \biggr. \ \ + \\
& \biggl. 8\pi r^2P(1+\Gamma) \biggr] -2\pi(E+P)r^2 \biggl[ \left( 1-e^{-\lambda} \right) \Gamma+\left(1+e^{-\lambda}\right)(2-\Gamma) \, \frac{1}{\gamma} \ \ + \biggr. \\
&\biggl. 8\pi Pr^2(1-\Gamma)\left(1+\frac{1}{\gamma}\right)\biggr],
\end{aligned}
\end{equation}

\begin{equation}
\begin{aligned}
\mathcal{S}_2 = & \biggl[ \pi r^2\left(1-\frac{1}{2}\,\Gamma\right)P(E+P)+\pi r^2\Gamma P^2  \ \ + \biggr. \\
&\frac{1}{16}\,\Gamma(E+P)\left(1-e^{-\lambda}\right)+\frac{1}{8}\left(E+P+\Gamma P \right) \left(1+7e^{-\lambda} \right) \ \ - \\
& \biggl. \frac{1}{8} \, r \, \Gamma' Pe^{-\lambda} \biggr],
\end{aligned}
\end{equation}
and
\begin{equation}
\begin{aligned}
\mathcal{S}_3 = & \Biggl[ -(E+P-\Gamma P)[\sigma^{(0)}]^2 r^2 e^{-\nu} \ \ + \\
&(E+P) \biggl[ \frac{31}{4} \, e^{-\lambda}-\frac{5}{2}-\frac{1}{4} \, e^{\lambda}+\frac{1}{2}\Gamma\left(e^{-\lambda}-1\right)\biggr] + \Gamma P \biggl[ -\frac{11}{4} \, e^{-\lambda} + \frac{3}{2} \ \ + \Biggr. \\
& \biggl. \frac{1}{4} \, e^{\lambda} \biggr] + 4\pi r^2(E+P)P \left( 3+e^{\lambda} \right) \left( \frac{1}{2}\Gamma -1 \right) + 4\pi r^2\Gamma P^2 \left( 1+e^{\lambda} \right) \ \ + \\
& \Biggl. \Biggl. 16\pi^2 r^4 P^2 e^{\lambda} \biggl[ (\Gamma-1)(E+P)+\Gamma P \biggr] + r \, \Gamma' \, P \, e^{-\lambda} \Biggr].
\end{aligned}
\end{equation}

\section{The Computations}
In this study, to compute nonrotating models (Section \ref{NRM}), rigid rotations (Section \ref{RR}), and spherical deformations (Section \ref{SD}), we use the corresponding numerical methods described in very detail in \citep{7GK08} (Sections 5.1 and 5.2). We then combine these methods with the numerical framework described in Section \ref{RO} for computing the zeroth- and second-order eigenfrequencies of pulsation. 

To implement all the required methods, and thus to compute the results presented here, we have written and used a $\mathit{Mathematica}^\circledR$ program.

\section{Numerical Results and Discussion}
\label{NRD}
To begin with, it is worth remarking here that, as it has been verified by several authors (see e.g. \citep{BWMB05}, Sections~4 and 7; see also \citep{PG03}, Section~5.3; for differentially rotating neutron stars, see \citep{7GK08}, Section~6), Hartle's perturbation method gives remarkably accurate results even when applied to rapidly rotating neutron stars, although this method has been developed as a slow-rotation perturbation method.

As discussed in \citep{6HF75} (Section~I), pulsars are identified as rotating neutron stars and, therefore, there is a strong interest in studying the influence of a rigid rotation on the properties of such relativistic objects. In particular, it is of great interest to compute the pulsation frequencies of the quasi-radial modes (i.e. modes which would be radial in the absence of rotation) for several models and thus to have a measure of the effect of general relativity on these frequencies. To that purpose, we have computed, and present in this section, relevant numerical results.      
  
Regarding our computations, we resolve four nonrotating general-relativistic polytropic models with polytropic indices $n=1.0, \, 1.5, \, 2.0, \, \mathrm{and} \, \, 2.5$. Each model is resolved for five central mass-energy densities: 
$E_\mathrm{c} = 10^{13}$, $3.16 \times 10^{13}$, $10^{14}$, $3.16 \times 10^{14}$, and  $10^{15}$ cgs. These values have been chosen to be below and relatively close to the ``maximum-mass densities'' of the corresponding models, being in fact the more interesting ones when considering neutron stars. It is worth mentioning here that the total mass $M$ of a model, treated as a function of the central density $E_\mathrm{c}$, $M = M(E_\mathrm{c})$, obtains a maximum value $M_\mathrm{max}$ for a specific value $E_\mathrm{c}^\mathrm{max}$; such a model is called ``maximum-mass model'', and the central density of this model is called ``maximum-mass density''. The maximum-mass densities of our models, computed by a method described in \citep{8GS11} (Section~4), are $E_\mathrm{c}^\mathrm{max} = 3.793 \times 10^{15}, \, 4.890 \times 10^{15}, \, 4.656 \times 10^{15}, \, \mathrm{and} \, \, 3.489 \times 10^{15}$ cgs, respectively (\citep{8GS11}, Tables~2, 3, 4, and 5, respectively). All models, studied here, have $E_\mathrm{c}^\mathrm{max} < 5 \times 10^{15}$ cgs. Accordingly, in our computations the sequence of central mass-energy densities is terminated at $E_\mathrm{c} = 10^{15}$ cgs. 

Next, each density case is resolved for the three lowest modes of pulsation: Mode~0, Mode~1, and Mode~2. For each mode, we compute a rigidly rotating configuration with angular velocity equal to the corresponding Keplerian angular velocity, $\Omega_\mathrm{K}$. Hartle's perturbation method uses proper expansions in the rotation parameter $\epsilon = \Omega/\Omega_\mathrm{max}$, where $\Omega_\mathrm{max} = \sqrt{G \, M / R^3}$ is the angular velocity for which mass shedding starts occuring at the equator of a star. Thus $\Omega_\mathrm{max}$ describes the Newtonian balance between centrifugal and gravitational forces. However, this Newtonian upper bound appears to be a rather overestimated limit for neutron stars. For such relativistic objects, the appropriate upper bound is $\Omega_\mathrm{K}$. Hence, if the angular velocity is slightly greater than $\Omega_\mathrm{K}$, then mass shedding occurs at the equator of a neutron star. $\Omega_\mathrm{K}$ can be computed by several methods (for a discussion on this matter, see \citep{PG03}; for a detailed description of  such a method, see \citep{BFGM05}; see also \citep{SG12}, and references therein, for results concerning general-relativistic polytropic models). In this study, the Keplerian angular velocities given in Tables \ref{ta1}, \ref{ta3}, \ref{ta5}, and \ref{ta7} have been computed by using the well-known ``Rotating Neutron Stars Package'' (RNS) \citep{S92}. 

Furthermore, we assume that the adiabatic indices $\Gamma$ (Equation~(\ref{Gamma})) and $\gamma$ (Equation~(\ref{gamma})) do coincide for the polytropic models under consideration, 
\begin{equation}
\Gamma = \gamma.
\end{equation}
In addition, a second step towards simplification is to assume, as several authors do (see e.g. \citep{6HF75}; for a different view, on the other hand, see e.g. \citep{14KR01}), that $\Gamma$ is associated with the polytropic index $n$ via the polytropic relation (cf. \citep{8GS11}, Equation~(5))  
\begin{equation}
\Gamma = 1 + \frac{1}{n}.
\label{GammaConst}
\end{equation}

In all tables of the present paper, parenthesized (positive or negative) integers following numerical values denote powers of ten. For example, the entry $3.16(13)$ is equal to $3.16 \times 10^{13}$ and the entry $3.51(-15)$ is equal to $3.51 \times 10^{-15}$.     

For the five density cases with polytropic index $n=1.5$, we can compare our results in Table~\ref{ta4} with respective results in Table~3 of \citep{6HF75}. To the purpose of such  comparisons, we have written Table~\ref{ta4} in exactly the same format with that of Table~3 of \citep{6HF75}. We find excellent agreement between respective results, except for two eigenfrequencies $[\sigma^2]^{(2)}$ belonging to Mode~2. The first one arises when $E_\mathrm{c} = 10^{14} \, \mathrm{cgs}$ and leads to a difference $\sim 4\%$ (our result is ``-15.95'', while the result in \citep{6HF75} is ``-16.6''); and the second one occurs when $E_\mathrm{c} = 10^{15} \, \mathrm{cgs}$ and leads to a difference $\sim 3\%$ (our result is ``-14.15'', while that in \citep{6HF75} is ``-13.7''). Since all other results do almost coincide, it seems that these two differences are of rather minor significance. 

Our main remarks on the numerical results presented in Tables \ref{ta4}, \ref{ta2}, \ref{ta6}, and \ref{ta8} have as follows. First, in all cases examined, the eigenfrequencies $[\sigma^2]^{(0)}$ and $[\sigma^2]^{(2)}$ increase in absolute value with the central density. Equivalently, since (for the cases examined) increasing central density implies increasing  gravitational mass, all the eigenfrequencies increase in absolute value with the gravitational mass. In addition, since increasing central density does also imply increasing Keplerian angular velocity for a rotating configuration, all the eigenfrequencies $[\sigma^2]^{(2)}$ increase in absolute value with the Keplerian angular velocity.

Second, all eigenfrequencies $[\sigma^2]^{(0)}$ are positive, thus representing stable nonrotating pulsating configurations. Likewise, the eigenfrequencies $[\sigma^2]^{(2)}$ are positive for Mode~0 and for the soft polytropic EOSs $n=2.5$ and $n=2.0$. Consequently, the effect of rotation is to stabilize such rotationally perturbed configurations.
On the other hand, the eigenfrequencies $[\sigma^2]^{(2)}$ are negative for Modes~1 and 2 of the soft EOSs $n=2.5$ and $n=2.0$, as well as for the three modes of the stiff EOSs $n=1.5$ and $n=1.0$, thus turning to destabilize the corresponding rotating configurations. It is worth mentioning here that, among the members of a collection of EOSs, the EOS deriving the larger value $P$ for a given $E$ is the ``stiffest'' EOS in the collection; while the EOS leading to the smaller value $P$ for the same $E$ is the ``softest'' EOS in this collection. Note that, for increasing polytropic index $n$, the polytropic EOSs are getting softer; equivalently, for decreasing $n$, the polytropic EOSs become stiffer.

Finally, in all cases examined, the zeroth-order eigenfrequencies $[\sigma^2]^{(0)}$ are $\sim$one order of magnitude greater than the respective squared Keplerian angular velocities $\Omega_\mathrm{K}^2$; this inequality is in fact a necessary condition for the  perturbation theory developed in \citep{13HTC72} and \citep{6HF75} to be valid, as discussed in detail in \citep{6HF75} (Section~II). Consequently, all the cases examined lie within the domain of applicability of the theory, and, especially, of Equation~\eqref{wdistort} for computing the second-order eigenfrequencies $[\sigma^2]^{(2)}$ which represent the rotationally induced changes in the pulsation eigenfrequencies. 

\clearpage

\begin{table} 
\begin{center}
\caption{Central mass-energy density and rest-mass density, gravitational mass, and radius of a nonrotating general-relativistic polytropic  model with polytropic index $n = 1.0$ and polytropic constant $K = 10^5 \, \mathrm{cgs} = 1.499 \times 10^{12} \, \mathrm{gu}$. The Keplerian angular velocities, appearing here, have been computed by RNS.\label{ta1}}
\begin{tabular}{ccccc} 
\hline\hline
$E_\mathrm{c}$ & $\varrho_\mathrm{c}$ & $M$ & $R$  & $\Omega_\mathrm{K}$\\
(cgs) & (cgs) & (gu)  & (gu) & (gu) \\ 
\hline
1.00(13) & 9.99(12) & 3.38(3) & 1.52(6) & 1.98(-8) \\ 
3.16(13) & 3.15(13) & 1.04(4) & 1.52(6) & 3.51(-8) \\
1.00(14) & 9.89(13) & 3.09(4) & 1.49(6) & 6.24(-8) \\
3.16(14) & 3.06(14) & 8.02(4) & 1.41(6) & 1.10(-7) \\ 
1.00(15) & 9.08(14) & 1.56(5) & 1.23(6) & 1.92(-7) \\
\hline
\end{tabular} 
\end{center}
\end{table}

\begin{table} 
\begin{center}
\caption{Central mass-energy density and rest-mass density, gravitational mass, and radius of a nonrotating general-relativistic polytropic  model with polytropic index $n = 1.5$ and polytropic constant $K = 5.380 \times 10^{15} \, \mathrm{cgs} = 3.389 \times 10^{7} \, \mathrm{gu}$. The Keplerian angular velocities, appearing here, have been computed by RNS.\label{ta3}}
\begin{tabular}{ccccc} 
\hline\hline
$E_\mathrm{c}$ & $\varrho_\mathrm{c}$ & $M$ & $R$  & $\Omega_\mathrm{K}$\\
(cgs) & (cgs) &      (gu)             & (gu)  & (gu) \\ 
\hline
1.00(13) & 9.96(12) & 1.58(4) & 3.13(6) & 1.42(-8) \\ 
3.16(13) & 3.13(13) & 2.73(4) & 2.56(6) & 2.52(-8) \\
1.00(14) & 9.80(13) & 4.54(4) & 2.08(6) & 4.46(-8) \\
3.16(14) & 3.04(14) & 7.08(4) & 1.66(6) & 7.81(-8) \\ 
1.00(15) & 9.22(14) & 9.84(4) & 1.29(6) & 1.35(-7) \\
\hline
\end{tabular} 
\end{center}
\end{table}

\begin{table}
\begin{center}
\caption{Central mass-energy density and rest-mass density, gravitational mass, and radius of a nonrotating general-relativistic polytropic  model with polytropic index $n = 2.0$ and polytropic constant $K = 10^{12} \, \mathrm{cgs} = 1.291 \times 10^{5} \, \mathrm{gu}$. The Keplerian angular velocities, appearing here, have been computed by RNS.\label{ta5}}
\begin{tabular}{ccccc} 
\hline\hline
$E_\mathrm{c}$ & $\varrho_\mathrm{c}$ & $M$ & $R$  & $\Omega_\mathrm{K}$ \\
(cgs) & (cgs) &      (gu)             & (gu) & (gu) \\ 
\hline
1.00(13) & 9.93(12) & 2.61(4) & 4.59(6) & 9.96(-9) \\ 
3.16(13) & 3.12(13) & 3.38(4) & 3.43(6) & 1.76(-8) \\
1.00(14) & 9.78(13) & 4.29(4) & 2.52(6) & 3.10(-8) \\
3.16(14) & 3.04(14) & 5.27(4) & 1.87(6) & 5.41(-8) \\ 
1.00(15) & 9.36(14) & 6.13(4) & 1.37(6) & 9.32(-8) \\
\hline
\end{tabular} 
\end{center}
\end{table}

\begin{table} 
\begin{center}
\caption{Central mass-energy density and rest-mass density, gravitational mass, and radius of a nonrotating general-relativistic polytropic  model with polytropic index $n = 2.5$ and polytropic constant $K = 1.500 \times 10^{13} \, \mathrm{cgs} = 2.980 \times 10^{3} \, \mathrm{gu}$. The Keplerian angular velocities, appearing here, have been computed by RNS.\label{ta7}}
\begin{tabular}{ccccc} 
\hline\hline
$E_\mathrm{c}$ & $\varrho_\mathrm{c}$ & $M$ & $R$  & $\Omega_\mathrm{K}$ \\
(cgs) & (cgs) &       (gu)            & (gu) & (gu) \\ 
\hline
1.00(13) & 9.93(12) & 1.96(4) & 5.31(6) & 6.74(-9) \\ 
3.16(13) & 3.13(13) & 2.16(4) & 3.76(6) & 1.19(-8) \\
1.00(14) & 9.84(13) & 2.36(4) & 2.55(6) & 2.10(-8) \\
3.16(14) & 3.08(14) & 2.54(4) & 1.82(6) & 3.67(-8) \\ 
1.00(15) & 9.60(14) & 2.68(4) & 1.29(6) & 6.38(-8) \\
\hline
\end{tabular} 
\end{center}
\end{table}

\begin{table}
\begin{center}
\caption{Eigenfrequencies of the lowest three modes of a nonrotating and a rotating general-relativistic polytropic model with polytropic index $n = 1.5$, and polytropic constant and Keplerian angular velocities as in Table~\ref{ta3}.\label{ta4}}
\begin{tabular}{ccccccr} 
\hline\hline
  & Mode 0 & & Mode 1 & & Mode 2 &  \\    
\hline
$E_c$ & $[\sigma^2]^{(0)}$ & $\frac{[\sigma^2]^{(2)}}{\Omega^2}$ & $[\sigma^2]^{(0)}$ & $\frac{[\sigma^2]^{(2)}}{\Omega^2}$ &  $[\sigma^2]^{(0)}$ & $\frac{[\sigma^2]^{(2)}}{\Omega^2}$ \\
(cgs) & (gu) & & (gu) & & (gu) \\ 
\hline
1.00(13) & 1.35(-15) & -0.33 & 6.36(-15) & -7.40 & 1.35(-14) & -17.27  \\ 
3.16(13) & 4.07(-15) & -0.32 & 1.96(-14) & -7.25 & 4.17(-14) & -17.17  \\
1.00(14) & 1.16(-14) & -0.31 & 5.87(-14) & -6.96 & 1.26(-13) & -15.95  \\
3.16(14) & 2.98(-14) & -0.28 & 1.66(-13) & -6.65 & 3.59(-13) & -15.51  \\ 
1.00(15) & 5.88(-14) & -0.22 & 4.27(-13) & -5.97 & 9.38(-13) & -14.15  \\
\hline
\end{tabular} 
\end{center}
\end{table}

\begin{table}
\begin{center}
\caption{Eigenfrequencies of the lowest three modes of a nonrotating and a rotating general-relativistic polytropic model with polytropic index $n = 1.0$, and polytropic constant and Keplerian angular velocities as in Table~\ref{ta1}.\label{ta2}}
\begin{tabular}{ccccccc} 
\hline\hline
  & Mode 0 & & Mode 1 & & Mode 2 &  \\  
\hline
$E_c$ & $[\sigma^2]^{(0)}$ & $[\sigma^2]^{(2)}$ & $[\sigma^2]^{(0)}$ & $[\sigma^2]^{(2)}$ &  $[\sigma^2]^{(0)}$ & $[\sigma^2]^{(2)}$ \\
(cgs) & (gu) & (gu) & (gu) & (gu) & (gu) & (gu) \\ 
\hline
1.00(13) & 3.51(-15) & -3.72(-16) & 1.50(-14) & -3.38(-15) & 3.19(-14) & -7.77(-15)  \\ 
3.16(13) & 1.08(-14) & -1.12(-15) & 4.66(-14) & -9.23(-15) & 9.94(-14) & -1.64(-14)  \\
1.00(14) & 3.18(-14) & -3.74(-15) & 1.41(-13) & -3.28(-14) & 3.01(-13) & -7.43(-14)  \\
3.16(14) & 7.99(-14) & -1.13(-14) & 3.87(-13) & -9.62(-14) & 8.37(-13) & -2.19(-13)  \\ 
1.00(15) & 1.33(-13) & -2.70(-14) & 8.68(-13) & -2.40(-13) & 1.92(-12) & -5.47(-13)  \\
\hline
\end{tabular} 
\end{center}
\end{table}

\begin{table}
\begin{center}
\caption{Eigenfrequencies of the lowest three modes of a nonrotating and a rotating general-relativistic polytropic model with polytropic index $n = 2.0$, and polytropic constant and Keplerian angular velocities as in Table~\ref{ta5}.\label{ta6}}
\begin{tabular}{ccccccc} 
\hline\hline
  & Mode 0 & & Mode 1 & & Mode 2 &  \\    
\hline
$E_c$ & $[\sigma^2]^{(0)}$ & $[\sigma^2]^{(2)}$ & $[\sigma^2]^{(0)}$ & $[\sigma^2]^{(2)}$ &  $[\sigma^2]^{(0)}$ & $[\sigma^2]^{(2)}$ \\
(cgs) & (gu) & (gu) & (gu) & (gu) & (gu) & (gu) \\ 
\hline
1.00(13) & 5.20(-16) & 4.07(-18) & 2.84(-15) & -7.02(-16) & 5.98(-15) & -1.69(-15)  \\ 
3.16(13) & 1.54(-15) & 1.44(-17) & 8.72(-15) & -2.16(-15) & 1.84(-14) & -5.21(-15)  \\
1.00(14) & 4.35(-15) & 6.89(-17) & 2.63(-14) & -6.02(-15) & 5.60(-14) & -1.15(-14)  \\
3.16(14) & 1.10(-14) & 2.64(-16) & 7.62(-14) & -1.85(-14) & 1.63(-13) & -4.07(-14)  \\ 
1.00(15) & 2.24(-14) & 1.13(-15) & 2.09(-13) & -5.25(-14) & 4.52(-13) & -1.24(-13)  \\
\hline
\end{tabular} 
\end{center}
\end{table}

\begin{table}
\begin{center}
\caption{Eigenfrequencies of the lowest three modes of a nonrotating and a rotating general-relativistic polytropic model with polytropic index $n = 2.5$, and polytropic constant and Keplerian angular velocities as in Table~\ref{ta7}.\label{ta8}}
\begin{tabular}{ccccccc} 
\hline\hline
  & Mode 0 & & Mode 1 & & Mode 2 &  \\    
\hline
$E_c$ & $[\sigma^2]^{(0)}$ & $[\sigma^2]^{(2)}$ & $[\sigma^2]^{(0)}$ & $[\sigma^2]^{(2)}$ &  $[\sigma^2]^{(0)}$ & $[\sigma^2]^{(2)}$ \\
(cgs) & (gu) & (gu) & (gu) & (gu) & (gu) & (gu) \\ 
\hline
1.00(13) & 1.62(-16) & 1.57(-17) & 1.25(-15) & -3.24(-16) & 2.59(-15) & -7.89(-16)  \\ 
3.16(13) & 4.72(-16) & 5.03(-17) & 3.85(-15) & -1.00(-15) & 8.02(-15) & -2.44(-15)  \\
1.00(14) & 1.31(-15) & 1.69(-16) & 1.19(-14) & -2.25(-15) & 2.59(-14) & -2.76(-15)  \\
3.16(14) & 3.23(-15) & 5.26(-16) & 3.54(-14) & -8.09(-15) & 7.53(-14) & -1.32(-14)  \\ 
1.00(15) & 6.16(-15) & 1.68(-15) & 1.03(-13) & -2.52(-14) & 2.18(-13) & -4.95(-14)  \\
\hline
\end{tabular} 
\end{center}
\end{table}


\clearpage

\begin{thebibliography}{01}

\bibitem[(1967)]{4H67} J. B. Hartle, ``Slowly Rotating Relativistic Stars. I. Equations of Structure,'' {\it The Astrophysical Journal}, Vol. 150, 1967, pp. 1005-1029.
doi:10.1086/149400

\bibitem[(1968)]{5HT68} J. B. Hartle and K. S. Thorne, ``Slowly Rotating Relativistic Stars. II. Models for Neutron Stars and Supermassive Stars,'' {\it The Astrophysical Journal}, Vol. 153, 1968, pp. 807-834. doi:10.1086/149707

\bibitem[(1972)]{13HTC72} J. B. Hartle, K. S. Thorne and S. M. Chitre, ``Slowly Rotating Relativistic Stars. VI. Stability of the Quasi-Radial Modes,'' {\it The Astrophysical Journal}, Vol. 176, 1972, pp. 177-194. doi:10.1086/151620

\bibitem[(1975)]{6HF75} J. B. Hartle and J. L. Friedman, ``Slowly Rotating Relativistic Stars. VIII. Frequencies of the Quasi-Radial Modes of an $n = 3/2$ Polytrope,'' {\it The Astrophysical Journal}, Vol. 196, 1975, pp. 653-660. doi:10.1086/153451

\bibitem[(2008)]{7GK08} V. S. Geroyannis and A. G. Katelouzos, ``Numerical Treatment of Hartle's Perturbation Method for Differentially Rotating Neutron Stars Simulated by General-Relativistic Polytropic Models,'' {\it International Journal of Modern Physics C}, Vol. 19, 2008, pp. 1863-1908. doi:10.1142/S0129183108013370

\bibitem[(2011)]{8GS11} V. S. Geroyannis and I. E. Sfaelos, ``Numerical Treatment of Rotating Neutron Stars Simulated by General-Relativistic Polytropic Models: A Complex-Plane Strategy,'' {\it International Journal of Modern Physics C}, Vol. 22, 2011, pp. 219-248.
doi:10.1142/S0129183111016269

\bibitem[(1994)]{9CST94} G. B. Cook, S. L. Shapiro and S. A. Teukolsky, ``Rapidly Rotating Polytropes in General Relativity,'' {\it The Astrophysical Journal}, Vol. 422, 1994, pp. 227-242. doi:10.1086/173721

\bibitem[(1965)]{10T65} R. F. Tooper, ``Adiabatic Fluid Spheres in General Relativity,'' {\it The Astrophysical Journal}, Vol. 142, 1965, pp. 1541-1562. doi:10.1086/148435

\bibitem[(2001)]{14KR01} K. D. Kokkotas and J. Ruoff, ``Radial Oscillations of Relativistic Stars,'' {\it Astronomy and Astrophysics}, Vol. 366, 2001, pp. 565-572.
doi:10.1051/0004-6361:20000216

\bibitem[(2005)]{BWMB05} E. Berti, F. White, A. Maniopoulou, and M. Bruni, ``Rotating Neutron Stars: An Invariant Comparison of Approximate and Numerical Space-Time Models,'' {\it Mon. Not. R. Astron. Soc.}, Vol. 358, 2005, pp. 923-938.
doi:10.1111/j.1365-2966.2005.08812.x

\bibitem[(2003)]{PG03} P. J. Papasotiriou and V. S. Geroyannis, ``A SCILAB Program for Computing General-Relativistic Models of Rotating Neutron Stars by Implementing Hartle's Perturbation Method,'' {\it International Journal of Modern Physics C}, Vol. 14, 2003, pp. 321-350. 10.1142/S0129183103004516

\bibitem[(2005)]{BFGM05} O. Benhar, V. Ferrari, L. Gualtieri and S. Marassi, ``Perturbative Approach to the Structure of Rapidly Rotating Neutron Stars,'' {\it Physical Review D}, Vol. 72, No. 4, 2005, Article ID: 044028. doi:10.1103/PhysRevD.72.044028

\bibitem[(2012)]{SG12} I. Sfaelos and V. Geroyannis, ``Third-Order Corrections and Mass-Shedding Limit of Rotating Neutron Stars Computed by a Complex-Plane Strategy,'' {\it International Journal of Astronomy and Astrophysics}, Vol. 2012,2, 2012, pp. 210-217. doi:10.4236/ijaa.2012.24027

\bibitem[(1992)]{S92} N. Stergioulas, ``Rotating Neutron Stars (RNS) Package,'' 1992. http:// www.gravity.phys.uwm.edu/rns/index.html. 

\end{thebibliography}
\end{document}